\documentclass{amsart}       

\usepackage{url}
\usepackage{graphicx}
\usepackage{textcomp}
\usepackage{amsmath}
\usepackage{amsfonts}
\usepackage{amssymb}
\usepackage{bbm}
\usepackage{graphicx}
\usepackage{subfig}
\usepackage{algorithm,wrapfig}
\usepackage{bm}
\usepackage{float}
\usepackage{lineno,multirow}
\usepackage{rotating}
\usepackage{adjustbox}

\begin{document}

\title[TDA on airflow for sleep stage annotation]{Topological Data Analysis Assisted Automated Sleep Stage Scoring Using Airflow Signals}

\author[qq1]{}
\author{Yu-Min Chung}
\address{Yu-Min Chung\\
AADS Eli Lilly and Company, Indianapolis, USA}
\email{yumchung@alumni.iu.edu}

\author{Whitney K. Huang}
\address{Whitney K. Huang\\
The School of Mathematical and Statistical Sciences,
Clemson University, SC, USA}
\email{wkhuang@g.clemson.edu}

\author{Hau-Tieng~Wu}
\address{Hau-Tieng Wu\\
Departments of Mathematics and Department of Statistical Science\\
Duke University, Durham, NC, United States of America}
\email{hauwu@math.duke.edu}

\begin{abstract}
{\bf Objective}: Breathing pattern variability (BPV), as a universal physiological feature, encodes rich health information. We aim to show that, a high-quality automatic sleep stage scoring based on a proper quantification of BPV extracting from the single airflow signal can be achieved.

{\bf Methods}: Topological data analysis (TDA) is applied to characterize BPV from the intrinsically nonstationary airflow signal, where the extracted features are used to train an automatic sleep stage scoring model using the XGBoost learner. The noise and artifacts commonly present in the airflow signal are recycled to enhance the performance of the trained system. The state-of-the-art approach is implemented for a comparison.

{\bf Results}: When applied to 30 whole night polysomnogram signals with standard annotations, the leave-one-subject-out cross-validation shows that the proposed features (overall accuracy 78.8\%$\pm$8.7\% and Cohen's kappa 0.56$\pm 0.15$) outperforms those considered in the state-of-the-art work (overall accuracy 75.0\%$\pm$9.6\% and Cohen's kappa 0.50$\pm 0.15$) when applied to automatically score wake, rapid eyeball movement (REM) and non-REM (NREM). The TDA features are shown to contain complementary information to the traditional features commonly used in the literature via examining the feature importance. The respiratory quality index is found to be essential in the trained system.

{\bf Conclusion}: The proposed TDA-assisted automatic annotation system can accurately distinguish wake, REM and NREM from the airflow signal.

{\bf Significance}: Since only one single airflow channel is needed and BPV is universal, the result suggests that the TDA-assisted signal processing has potential to be applied to other biomedical signals and homecare problems other than the sleep stage annotation.

{\bf Keywords:} sleep stage annotation, topological data analysis, breathing pattern variability,  XGBoost, respiratory quality index

\end{abstract}

\maketitle

\section{Introduction}
\label{sec:intro}

Sleep is an integral aspect of human health \cite{kryger2017principles}, characterized by distinct patterns of brain activity discernible through electroencephalogram (EEG) measurements. Although much remains unknown, numerous connections between various sleep stages and health conditions have been established \cite{karni1994dependence,kang2009amyloid}. Additionally, it is widely recognized that many public disasters are linked to poor sleep conditions \cite{colten2006functional}. Clearly, an efficient monitoring of sleep dynamics could serve as the initial step in improving health quality and mitigating public disasters. The gold standard of studying the sleep dynamics is through quantifying various EEG patterns, where sleep is classified into several stages, including wakefulness, rapid eye movement (REM), N1, N2, and N3 \cite{AASM2020}. Typically, N1 and N2 are grouped together as light sleep, while N3 is known as deep sleep. 
From a physiological standpoint, sleep encompasses not only brain activity, but also affects every facet of the body \cite{kryger2017principles}, which enables the possibility to estimate sleep stage from non-EEG physiological signals.
There have been many papers discussing how to use non-EEG channels to automatically annotate sleep stages. Among many non-EEG channels, researchers usually consider electrocardiogram, respiratory signal, photoplethysmogram, actinogram, and sound. See, for example, \cite{matar2018unobtrusive,imtiaz2021systematic}, for a systematic review. Among those works that involves respiratory signal, most of them combine other channels, e.g., body movement \cite{willemen2013evaluation,hwang2016unconstrained, gaiduk2021estimation,dietz2021proof}, heart rate and body movement \cite{gaiduk2018automatic,stuburic2020deep}, heart rate \cite{fonseca2015sleep,bakker2021estimating}, to achieve the goal. To our knowledge, only few results focus on using only the respiratory signal \cite{sloboda2011simple,long2014analyzing,long2014measuring,tataraidze2015sleep,yang2016sleep}, where the authors mainly focus on features in the time domain and frequency domain.

\begin{figure*}
\includegraphics[width=0.48\linewidth]{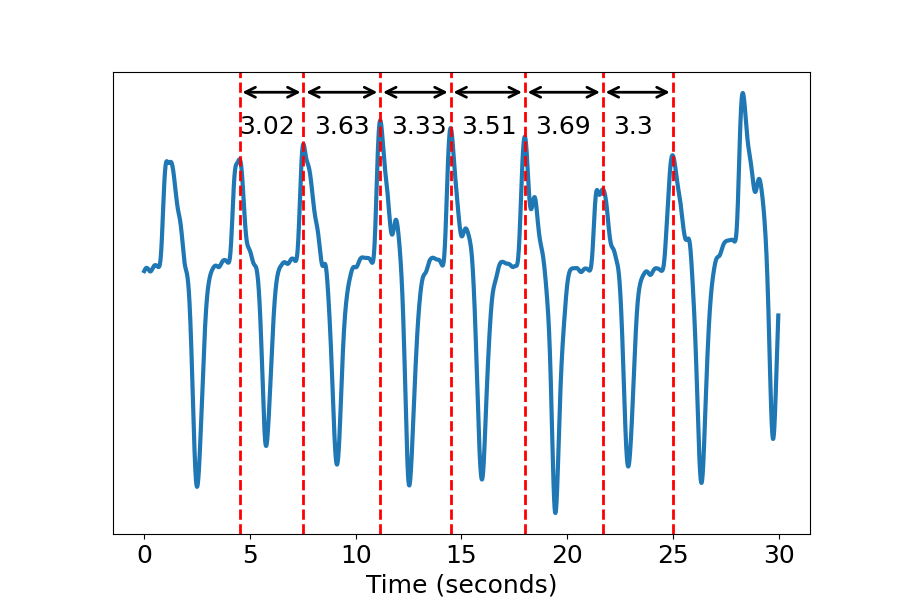}
\includegraphics[width=0.48\linewidth]{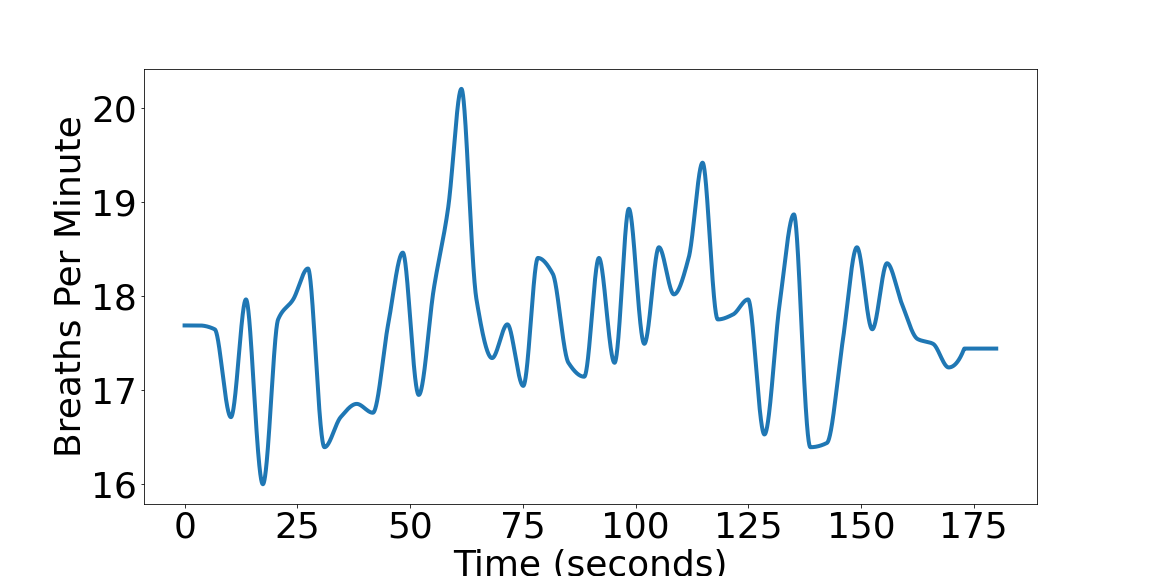}
\caption{An illustration of airflow signal and the associated instantaneous respiratory rate (IRR), where an 180-second airflow signal is zoomed in and shown in (a), and its IRR is shown in (b). \label{figure1:BPV}}
\end{figure*}

The potential to estimate sleep stage from the respiratory signal is based on the intimate relationship between sleep stage and breathing pattern variability (BPV) \cite[Chapter 4]{kryger2017principles}. In short, the body's requirement for oxygen and removal of carbon dioxide fluctuates depending on the stage of sleep. During wakefulness and the initial stages of sleep, breathing is characterized by irregular breathing patterns, shallow breathing, and low amplitude of respiratory muscle activity. As sleep progresses to deeper stages, breathing becomes more regular, and respiratory muscle activity increases. During REM sleep, which is associated with dreaming, breathing becomes even more irregular and shallow. The variability in breathing during REM sleep is thought to be due to changes in the activity of the respiratory centers in the brainstem, which regulate breathing. See Figure \ref{figure1:BPV} for an illustration of BPV. We therefore hypothesize that a proper quantification of BPV would allow us to design a more accurate automatic sleep stage scoring system.

Topological data analysis (TDA), a modern data analysis framework motivated by analyzing complicated high dimensional datasets, has gained a significant amount of attention as a versatile tool for data exploration and analysis \cite{carlsson2009topology,edelsbrunner2010computational,carlsson_vejdemo-johansson_2021}. Its theoretical foundation has been actively established under the framework of algebraic topology. The basic idea underlying TDA is that the number of holes of different dimensions (also known as the Betti numbers), under a proper mathematical definition, characterizes how the data is organized. TDA has demonstrated its potential for analyzing complex datasets from diverse fields such as biology, physics, neuroscience, and social science, among others (see e.g. \cite{carlsson_vejdemo-johansson_2021}), and is an area of active research, with new algorithms and applications being developed regularly. For example, the popular Mapper algorithm \cite{singh2007topological}, which provides a topological representation of the data, and the persistent homology algorithm \cite{carlsson2009persistent}, which captures topological features across different scales. Time series, as a ubiquitous data type, has been proven to benefit substantially from this simple but powerful TDA idea. For example, researchers have applied it to study voices, body motions \cite{seversky2016time,venkataraman2016persistent}, the trading records \cite{gidea2018topological}, electroencephalogram (EEG) signals \cite{piangerelli2018topological,wang2018topological}, instantaneous heart rate \cite{chung2021persistent}, among many others. See \cite{skaf2022topological} for a current review for its application in biomedical field. Recall that in the traditional time series analysis, unless a detailed mathematical model is possible, the majority of analysis tools are based on time or frequency domain analysis. However, biomedical time series is usually nonstationary and challenging to model precisely, which might limit the applicability of these analysis tools. The respiratory signal of interest in this study is a typical example. To handle such challenges, we consider TDA as an alternative tool that is capable of characterizing nonstationary time series from the perspective different from the traditional approaches. 

In this paper, motivated by the aforementioned physiological connection between BPV and sleep stage, the need for an efficient homecare monitoring of sleep dynamics, and the potential of TDA to quantify BPV from the nonstationary respiratory signal, we design an automatic sleep stage scoring system based on TDA features of the airflow signal. We illustrate its usefulness by analyzing a real world polysomnogram (PSG) database, and comparing our method with existing state-of-the-art algorithms. The paper is organized in the following. In Section \ref{subsec:tda}, we briefly summarize TDA and how to apply it to time series to extract useful features. We also provide reference for the interested readers in more details. In Section \ref{sec:sleep stages}, the designed features and learning model are detailed, and the results of analyzing a polysomnogram database is shown. In Section \ref{sec6}, we close the paper with a discussion and conclusion.

\section{Topological Data Analysis}\label{subsec:tda}

TDA is an emerging framework of data analysis. Recently, it has been widely applied to study nonlinear and nonstationary time series, for example \cite{seversky2016time,venkataraman2016persistent,gidea2018topological,piangerelli2018topological,wang2018topological,chung2021persistent,skaf2022topological}.  
A general pipeline for applying TDA tools to data can be found in Figure 2 in \cite{chung2021persistent}.  It is originated from the field called algebraic topology, and has recently gained attention due to its success in several applications.  TDA consists of a suite of tools aims to provide a means to characterize the ``shape'' of data. Persistent homology, one of main TDA tools, can be thought as a multi-scale analysis tool. 
We refer readers with interest to \cite{edelsbrunner2010computational} for necessary theoretical foundation of algebraic topology and \cite{chung2021persistent} for a quick lecture of necessary information.

\subsection{Filtration and persistence diagram for time series analysis}
To apply TDA, we need two main objects, a filtration of the dataset of interest and its corresponding persistence diagram (PD).  
A filtration is a sequence of nested subsets of the data, where each subset corresponds to a different threshold value for a chosen function on the data. A specific construction of the filtration depends on the application. The PD records the birth and death times of topological features encoded in a filtration, such as connected components, holes, and voids, as the filtration parameter varies. Specifically, PD represents each topological feature as a point in the plane, where the x-coordinate corresponds to the birth time of the feature, and the y-coordinate corresponds to the death time of the feature. Features that persist for a long time in the filtration are represented as points far away from the diagonal line, while short-lived features are represented as points near the diagonal.  {\em Sublevel set filtration} and {\em Rips complex filtration} and their associated PDs are commonly applied in the time series analysis (see e.g \cite{seversky2016time,venkataraman2016persistent,gidea2018topological,piangerelli2018topological,wang2018topological,chung2021persistent,skaf2022topological}). For more rigorous treatment on filtration and PD, we refer readers to \cite{edelsbrunner2010computational,carlsson_vejdemo-johansson_2021}.  Below we follow the work \cite{chung2021persistent} and describe sublevel set, Rips complex filtration, and their PDs in the time series setup.

\begin{figure*}[htb!]
	\subfloat[]{\includegraphics[width=0.33\linewidth]{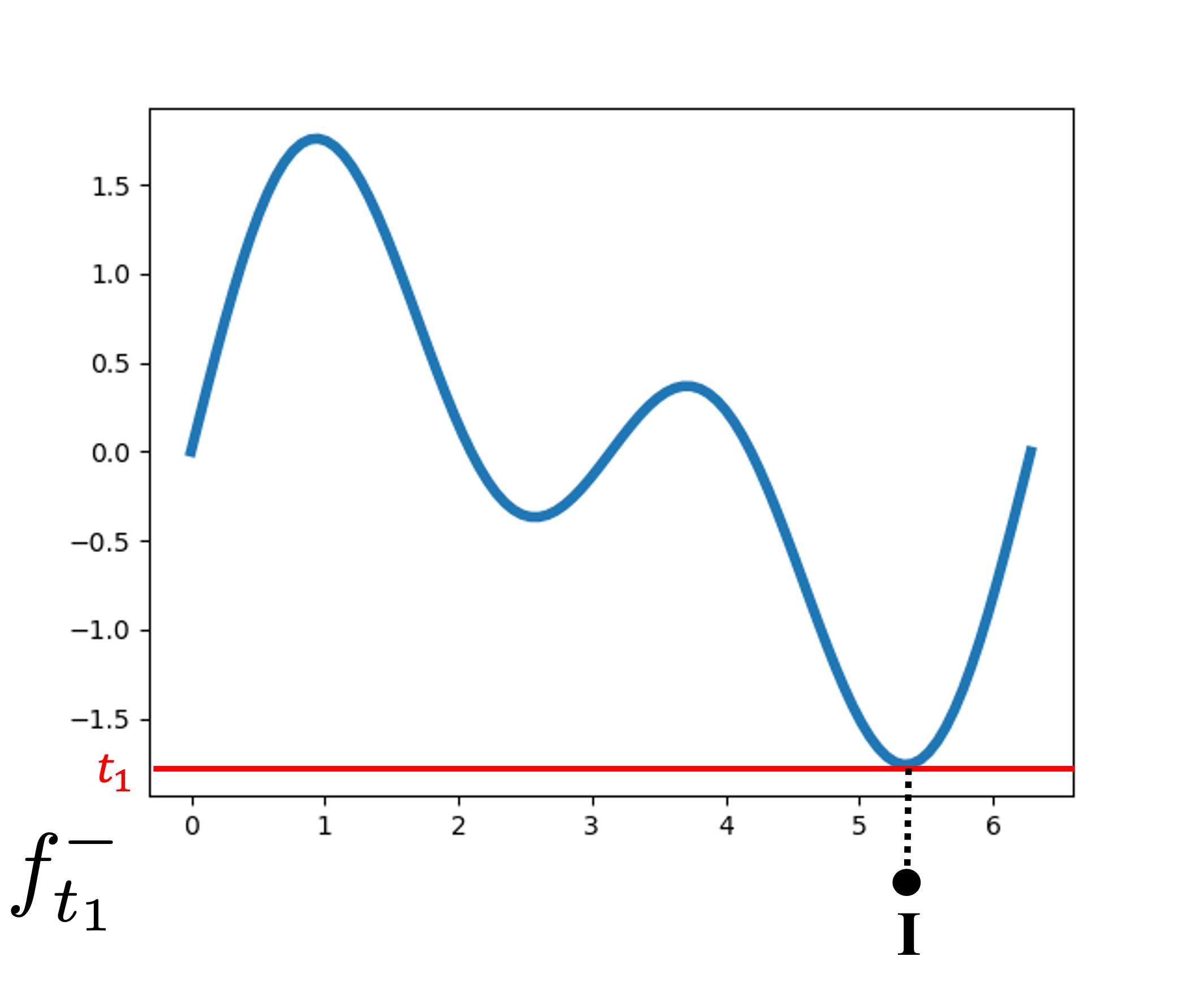}}
	\subfloat[]{\includegraphics[width=0.33\linewidth]{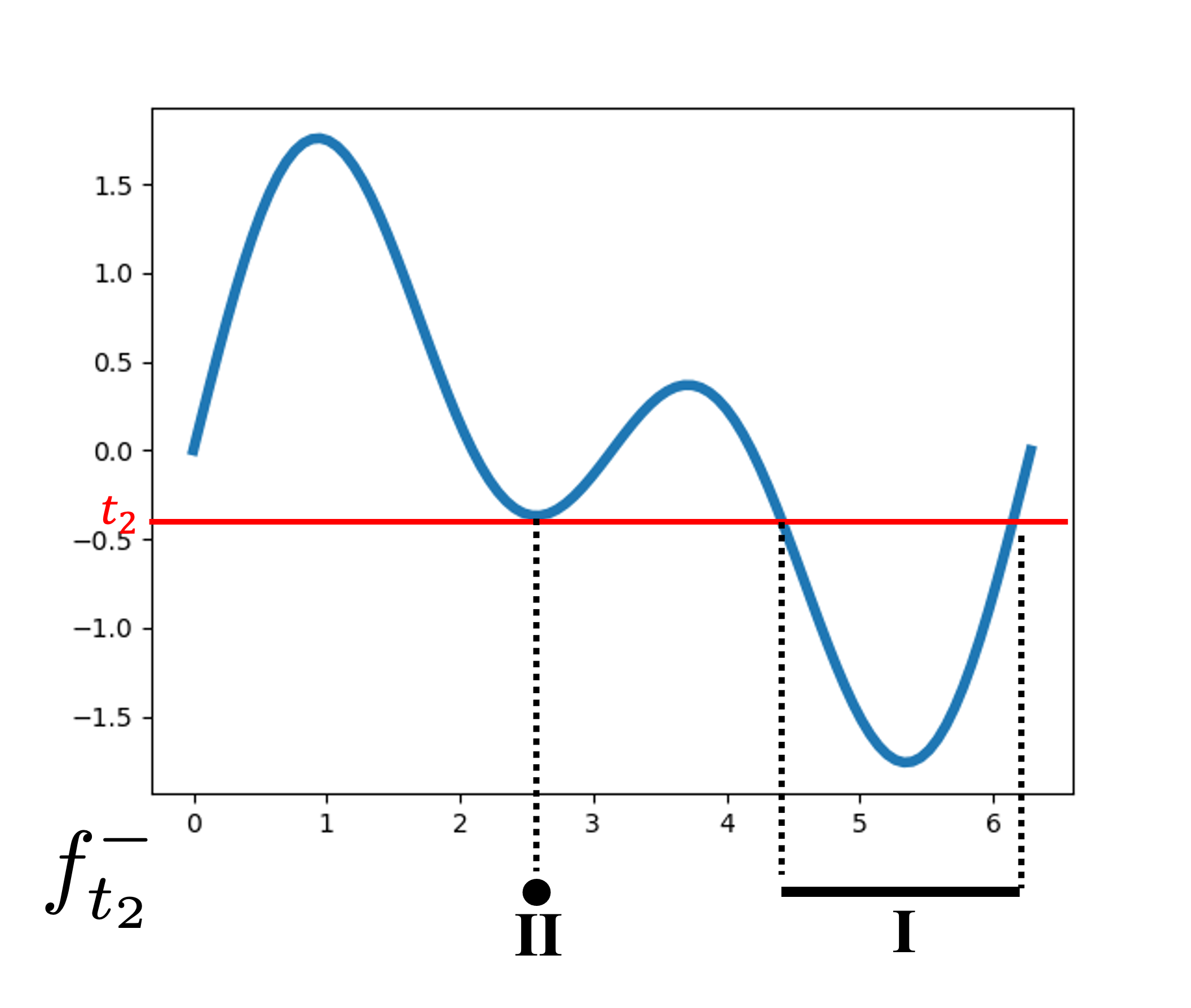}}
	\subfloat[]{\includegraphics[width=0.33\linewidth]{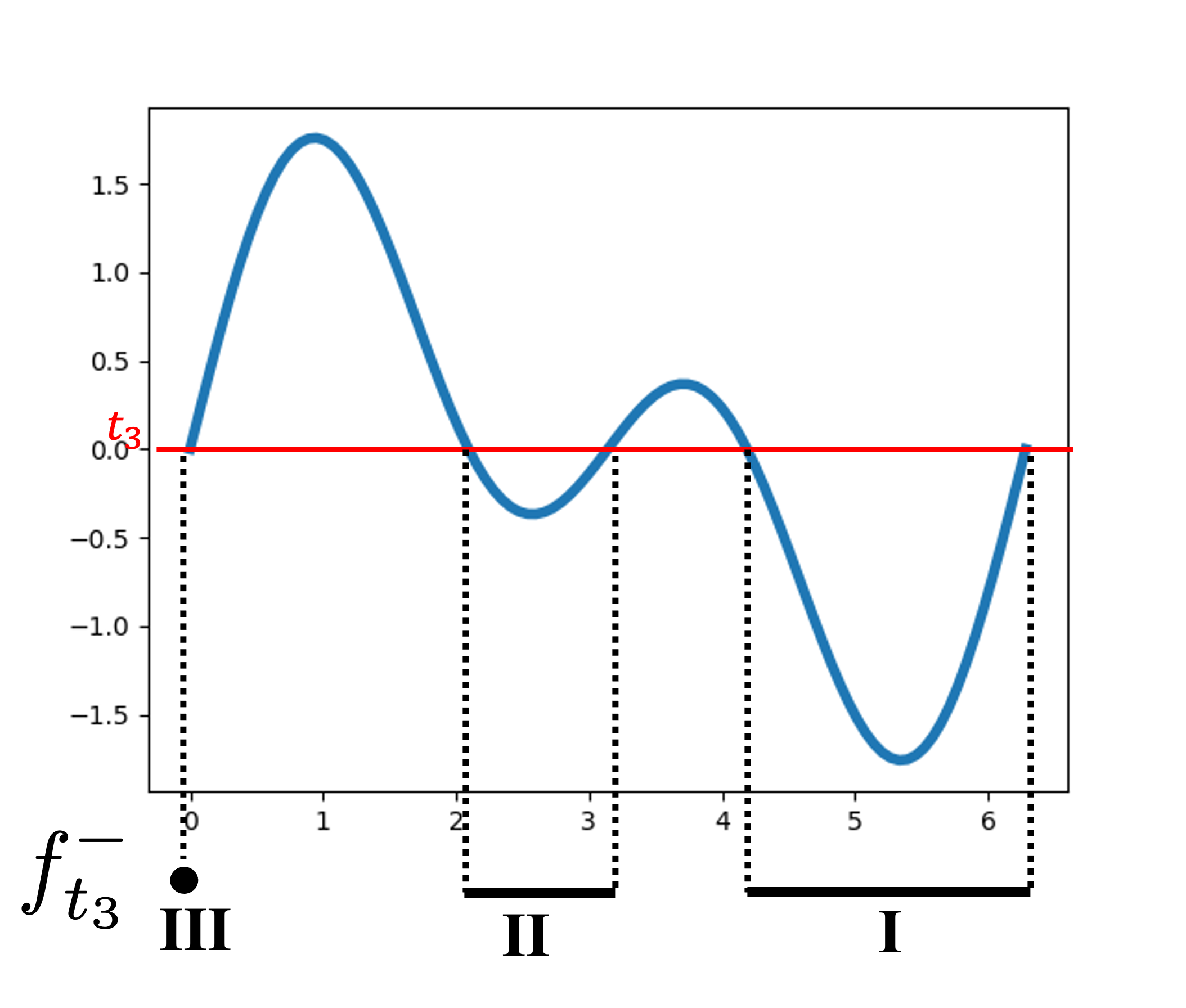}}\\
	\subfloat[]{\includegraphics[width=0.33\linewidth]{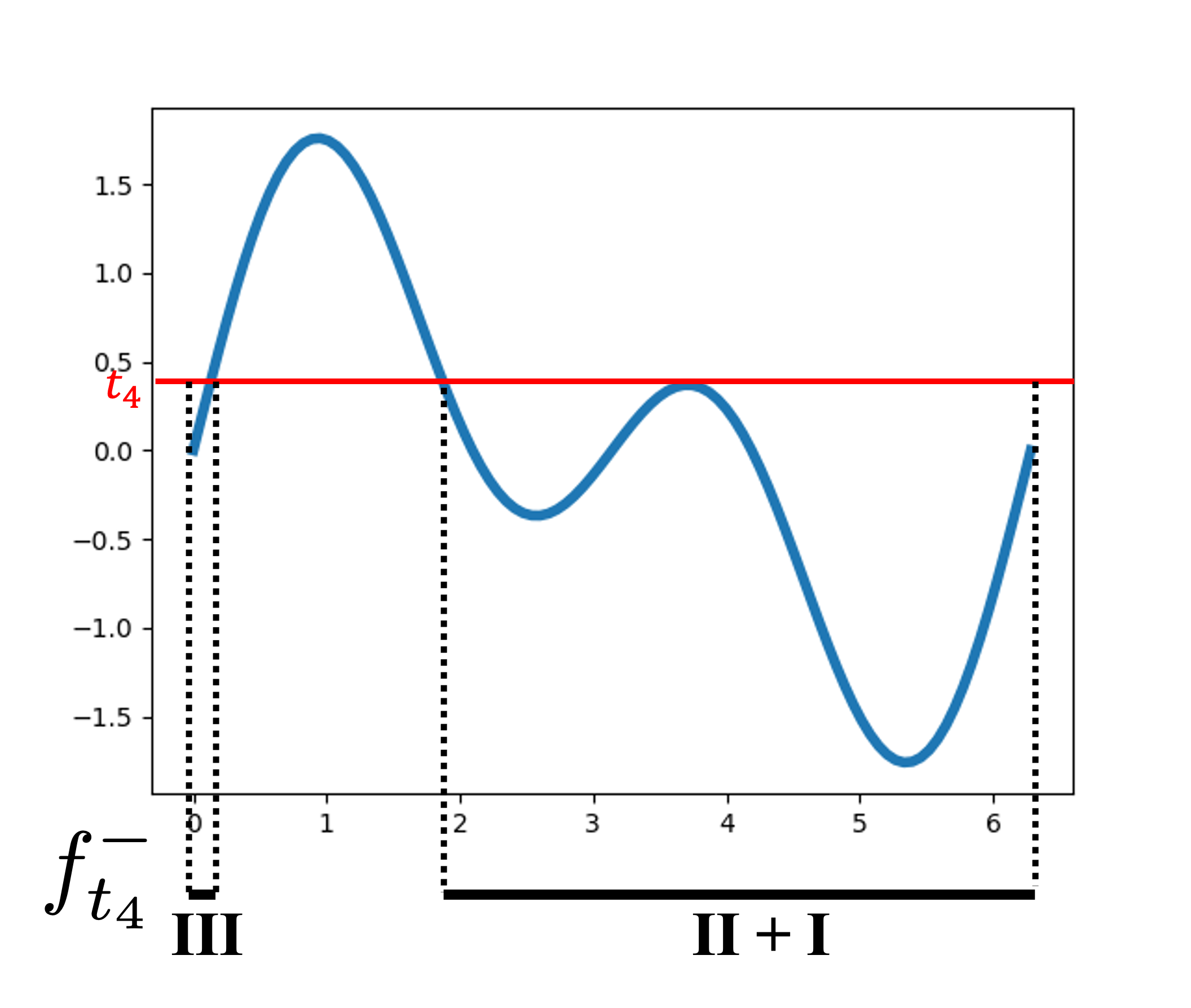}}
	\subfloat[]{\includegraphics[width=0.33\linewidth]{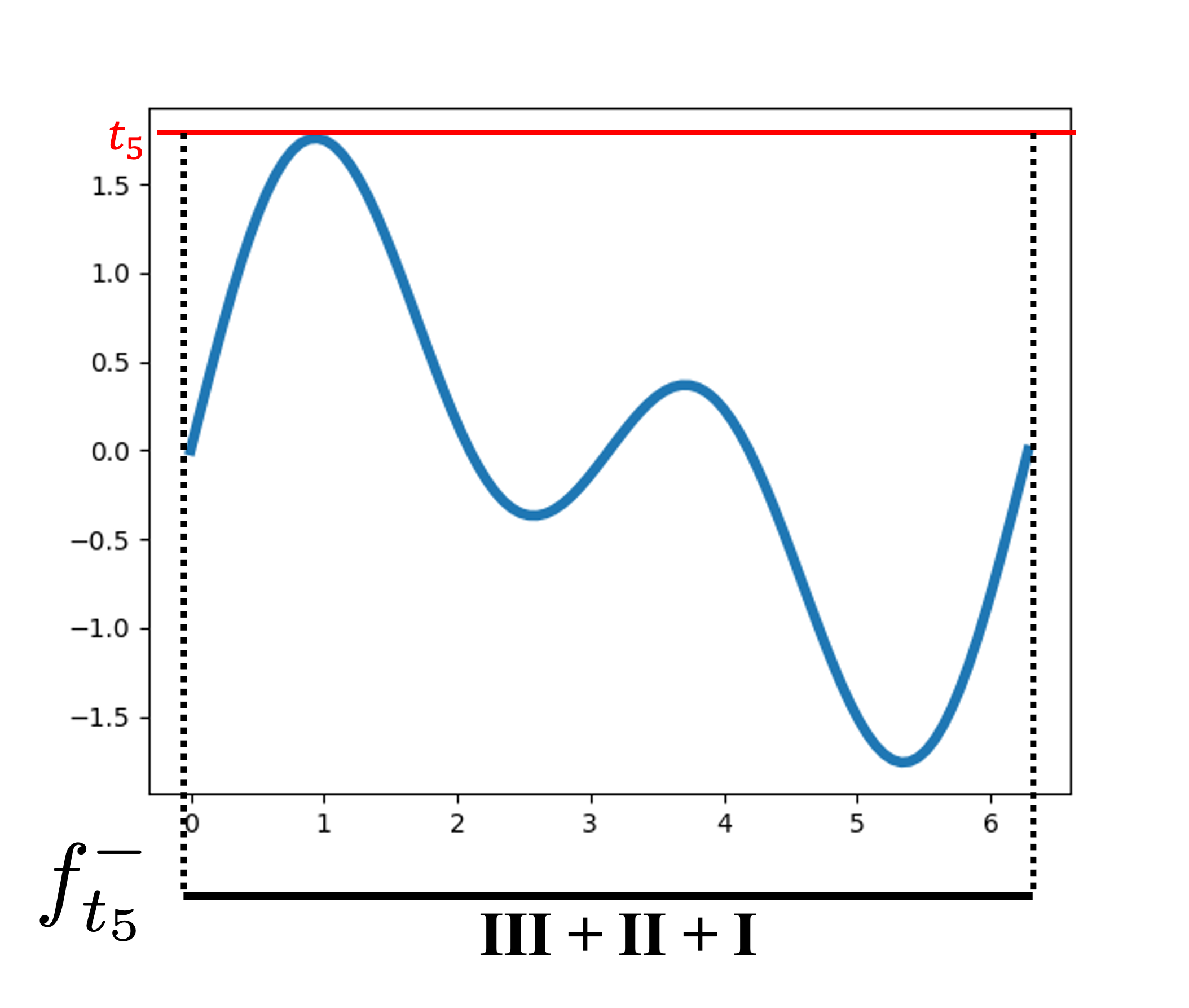}}			
	\subfloat[]{\includegraphics[width=0.37\linewidth]{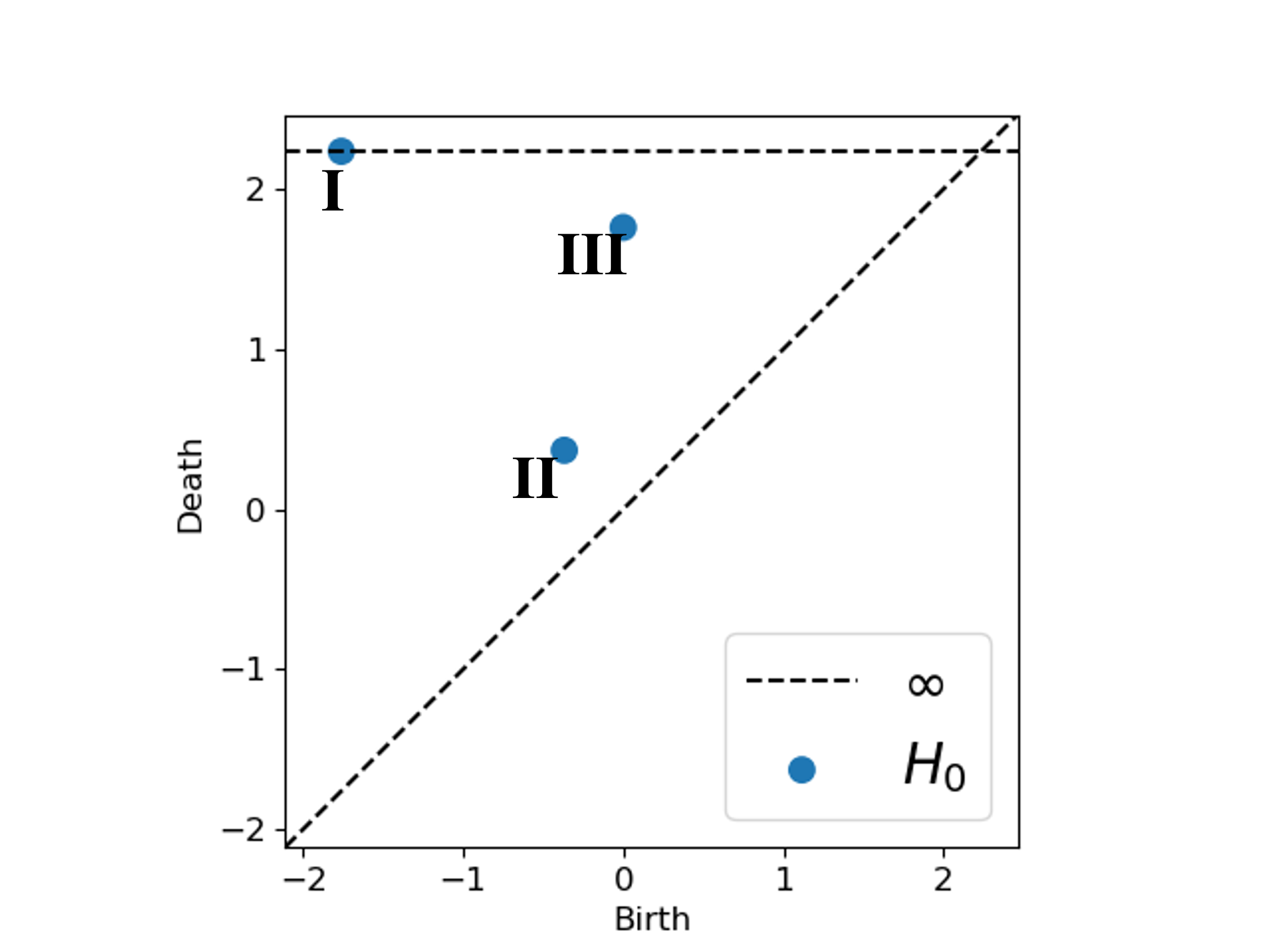}}				
	\caption{Illustration of sublevel set filtration.  Blue curve is the time series sampled from $\sin(x)+\sin(2x)$ for $x\in [0,2\pi]$. Red horizontal line indicates the current threshold value.  The  black sets labeled as roman number I, II, and III below each subplot demonstrate the sublevel sets. (a)-(e) Illustration of sublevel sets, $f_{t_i}$, for $t_1,~t_2,~t_3,~t_4,~t_5$, respectively. (f) The corresponding persistence diagram.} 
	\label{fig:sublevel set filtration}
\end{figure*}
 
The sublevel set filtration of a given continuous function $f$ is defined in the following way. Let $t\in \mathbb{R}$ be a threshold value.  The sublevel set of $f$ at $t$ is defined as 
\[
f_t^-:=\{x| f(x)\leq t\}\,.
\]  
One can show that $f_{t_1}^- \subseteq f_{t_2}^-$ when $t_1 \leq t_2$. Thus, $\{f_{t}^-\}_{t\in \mathbb{R}}$ form a filtration. The sublevel set filtration of the given time series contains information about the local minimum and local maximum and is related to the Morse theory \cite{matsumoto2002introduction}. For example, Figure~\ref{fig:sublevel set filtration}(a)-(e) demonstrate the sublevel sets for the function $f(x) = \sin(x) + \sin(2x)$ on $x\in [0,2\pi]$, and the corresponding PD is shown in (f).  The threshold values, $t_i$, are shown as red horizontal lines in Figure~\ref{fig:sublevel set filtration}. Intuitively, sublevel sets are the regions that are below the red horizontal lines. We start the filtration with the smallest value of the function as shown in Figure~\ref{fig:sublevel set filtration} (a) and we notice that the sublevel set, $f^-_{t_1}$, is a set of single point as depicted as the Roman number I.  As shown in Figure~\ref{fig:sublevel set filtration}(b)-(e) we observe that the sublevel set increases as the threshold value increases.   The corresponding PD shown in Figure~\ref{fig:sublevel set filtration}(f) records information about changes of these sublevel sets.  At $t_1$, the component I first appears, so we associate I with the {\it birth} value $t_1$. Similarly, the components II and III first appear at $t_2$ and $t_3$, respectively, so we associate them with the birth value $t_2$ and $t_3$, respectively.  At $t_4$, the components I and II merge, and we associate II with the {\it death} value $t_4$ and I continues to live (such choice is commonly known as the {\em elder rule} \cite{edelsbrunner2010computational}).  At $t_5$, the components III and I+II merge, so we associate III with the death value $t_5$ and since $t_5$ is the largest value of the function, the filtration ends here.  We note that the component I continues to live even though the filtration ends.  By convention (see e.g. \cite{edelsbrunner2010computational}), the death value of $\infty$ is assigned to the component I.  Therefore, the PD shown in Figure~\ref{fig:sublevel set filtration} contains three points: $(t_1,  \infty)$, $(t_2, t_4)$, and $(t_3, t_5)$. 
Given a time series signal $f$, we denote the sublevel set persistence diagram by $D_0^{\texttt{Sub}}(f)$. Here, the index $0$ means the 0-dimensional hole or disjoint connected components.  One can consider higher dimensional holes, but since the sublevel set is a 1-dimensional object, higher dimensional holes are trivial (see e.g. \cite{carlsson_vejdemo-johansson_2021}).  

Next, we describe the Rips complex filtration. This filtration depends on a widely applied nonlinear time series analysis tool, the Takens' embedding \cite{takens2006detecting}. The Takens' embedding converts a time series into a high dimensional point clouds, and under proper and mild assumptions, it allows users to recover the underlying phase space of the observed dynamics. 
Given a time series $f_n$, where $n=1,\ldots,N$, pick an appropriate time delay $\tau\in \mathbb{N}$ and embedding dimension $d\in \mathbb{N}$. 
Construct a set of $d$-dimensional vectors $x_n$ by taking $d$ consecutive samples of the time series with a time lag of $\tau$, i.e., $x_n:=[f_n, f_{n+\tau}, f_{n+2\tau},\ldots, f_{n+(d-1)\tau}]^\top$. The set of vectors, $x_1,\ldots, x_{N-(d-1)\tau}$, is called the Takens' embedding of the time series.
Note that $\tau$ should be chosen to be a characteristic time scale of the system being studied, while $d$ should be large enough to capture the intrinsic geometry of the system but not too large to avoid the complexity of the embedding. For example, Figure \ref{fig:demo rips pd resp} (g), (h), and (i) demonstrate the Takens' embedding of (a), (b), and (c), respectively, where the parameter $\tau=100$ and $d=3$.
\begin{figure*}[htb!]
	\subfloat[point cloud]{\includegraphics[width=0.25\linewidth]{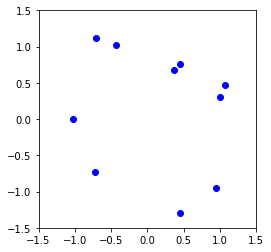}}
	\subfloat[$\epsilon_1$]{\includegraphics[width=0.25\linewidth]{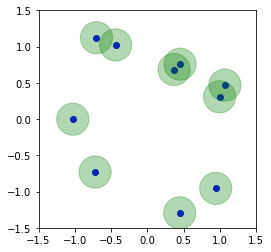}}
	\subfloat[$\epsilon_2$]{\includegraphics[width=0.25\linewidth]{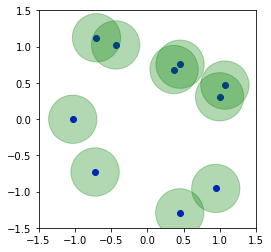}}
	\subfloat[$\epsilon_3$]{\includegraphics[width=0.25\linewidth]{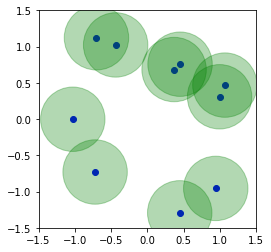}}\\
	\subfloat[$\epsilon_4$]{\includegraphics[width=0.25\linewidth]{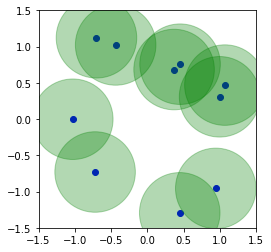}}
	\subfloat[$\epsilon_5$]{\includegraphics[width=0.25\linewidth]{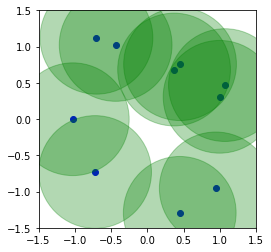}}				
	\subfloat[$\epsilon_6$]{\includegraphics[width=0.25\linewidth]{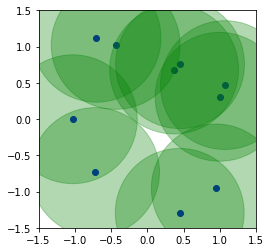}}				
	\subfloat[$\epsilon_7$]{\includegraphics[width=0.25\linewidth]{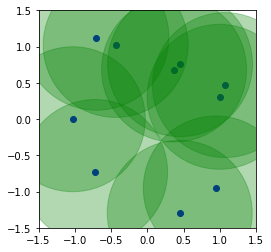}}					
	\caption{Illustration of Rips complex filtration. (a) Sample point cloud. (b)-(h) Illustration of Rips complexes at various radius, where $\epsilon_1<\epsilon_2<\ldots<\epsilon_7$.}
	\label{fig:rips complex filtration}
\end{figure*}
We refer readers with interest to \cite{takens2006detecting} for theoretical justification. Once we obtain a high dimensional point clouds, the Rips complexes is constructed by the following steps. For each distance threshold value $\epsilon>0$, determine a ball of radius $\epsilon$ around each point in the dataset. Then, construct a Rips complex, which is a simplicial complex whose simplices are formed by the union of balls of radius $\epsilon$ centered at the points in the dataset. For more technical and rigorous treatment about simplicial complexes, we refer interested readers to the classic text \cite{hatcher2002algebraic}. Conceptually, we may think of Rip complex with radius $\epsilon$ as being simply the union of balls of radius $\epsilon$ centered at the points in the dataset.  For instance, Figure~\ref{fig:rips complex filtration}(a) shows a sample point cloud and (b)-(h) demonstrate the Rip complex with various radius. Clearly, the Rips complex grows in size as $\epsilon$ increases and new simplices are added. As a result, each Rips complex associated with a larger $\epsilon$ contains all the simplices of the smaller $\epsilon$. This nested relationship leads to a filtration of Rips complexes indexed by $\epsilon$. See Figure~\ref{fig:rips complex filtration} for an example.  Figure~\ref{fig:rips complex filtration}(a) is the given point cloud that consists of 10 points. In Figure~\ref{fig:rips complex filtration}(b), there are 7 isolated regions in the Rips complex associated with the radius $\epsilon_1$. These isolated regions are called the {\em 0-dimensional holes}.  As radius increases, more regions merge.  In Figure~\ref{fig:rips complex filtration}(c), two balls at the lower right corner merge and four balls at the upper right corner merge, so there are now only 5 isolated regions.  One may observe the process in (d)-(h).  In particular, in Figure~\ref{fig:rips complex filtration} (f), all the balls merge and they form a hole in the middle (white region enclosed by green circles).  This hole is known as the {\em 1-dimensional hole}. In Figure~\ref{fig:rips complex filtration} (h), the 1-dimensional hole is filled.  Similar to the discussion in the sublevel set PD, one may consider the PD of the Rips complex filtration. The main difference is that there will be two corresponding PDs: one for 0-dimensional holes, and another for 1-dimensional holes. To sum up, given a time series signal $f$, we first construct the Takens' embedding of $f$ using the parameter $(\tau, d)$, then consider the Rips complex filtration on the point cloud of Takens' embedding, and lastly compute its PDs.  For simplicity, we denote such Rips complex persistence diagram by $D_i^{\texttt{Rip}}(f)$ for $i=0,~1$.  Here, the Takens' embedding of $f$ is a $d$-dimensional object.  In addition to the 0-dimensional holes, we also consider the 1-dimensional holes.
The PD of Rips complex filtration of a time series contains information other than that hidden in the PD of sublevel set filtration, for example, the periodicity \cite[Section 3.2, p. 61]{edelsbrunner2010computational}. 

To motivate our study, Figure~\ref{fig:demo rips pd resp}(a), (b) and (c) show examples of respiratory signals during different sleep stages Awake, REM, and NREM, respectively and their corresponding IRRs are shown in Figure~\ref{fig:demo rips pd irr} (a), (b) and (c), respectively.   Figure~\ref{fig:demo rips pd resp}(d), (e), and (f) are the sublevel set PD for the respiratory signals (a), (b), and (c), respectively.  Similarly,   Figure  \ref{fig:demo rips pd irr} (d), (e), and (f) are the sublevel set PD for the IRR signals (a), (b), and (c), respectively.  Figure~\ref{fig:demo rips pd resp} (j), (k), and (l) are the Rips complex PDs for the respiratory signals (a), (b), and (c), respectively, where blue points indicate the 0-dimensional holes, and orange points indicate 1-dimensional holes. We observe from Figures \ref{fig:demo rips pd resp} and \ref{fig:demo rips pd irr} that visually PDs of different signals at different sleep stages are different. To further quantify such differences, the next subsection devotes to the topics of summarizing the PDs.

\begin{figure*}[hbt!]
\centering
\includegraphics[trim=0 30 0 50,width=.95\linewidth]{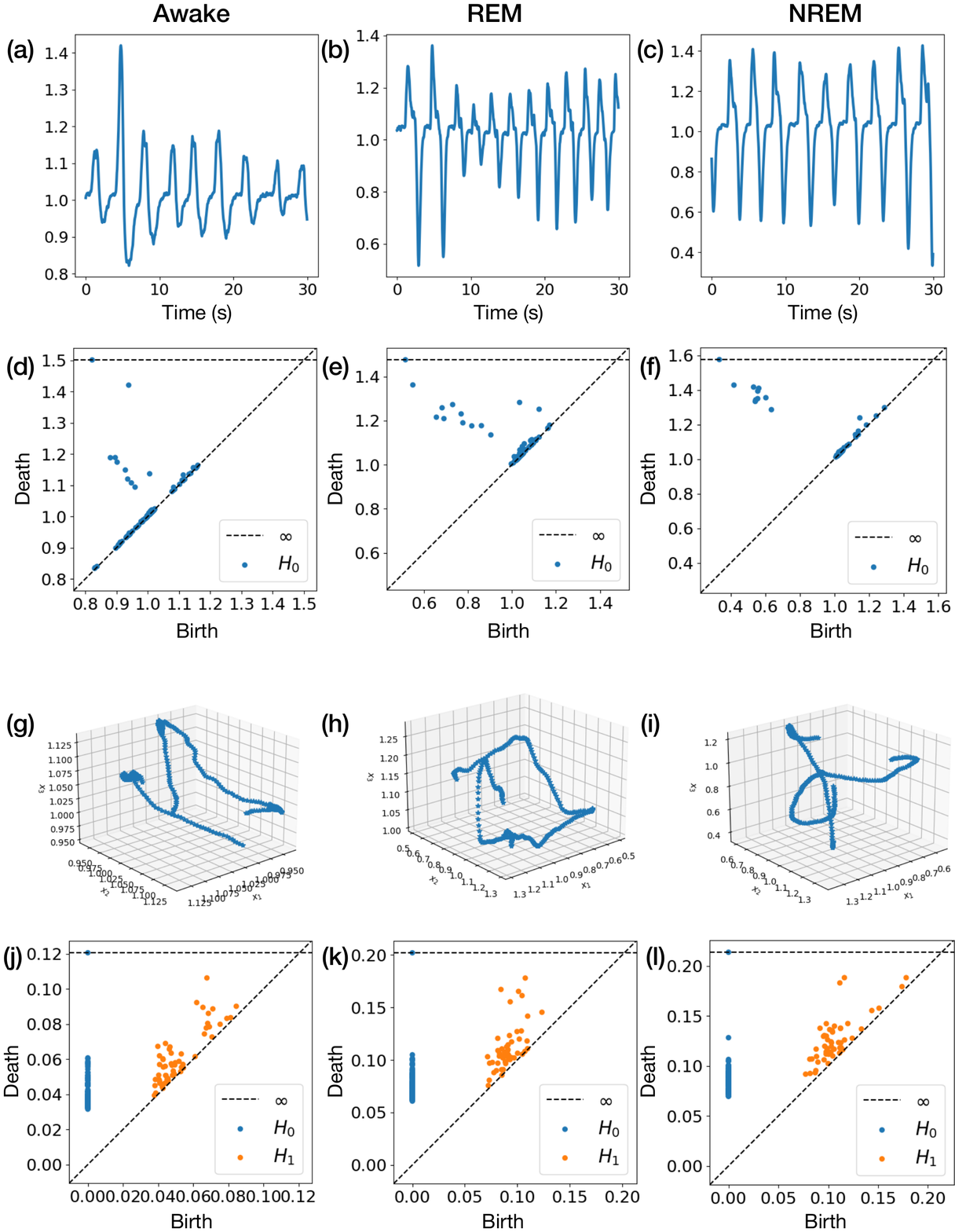}
\caption{Examples of respiratory signals and their persistence diagrams during different sleep stages. (a)-(c) Respiratory signals recorded during Wake, REM, and NREM sleep stages, respectively.  (d)-(f) Sublevel set persistence diagrams of (a)-(c), respectively. (g)-(i) Takens' embedding of (a)-(c), respectively.  The embedding parameter is $3$ and the delay parameter is $1$ second. (j)-(l) Rips complex persistence diagrams of (g)-(i), respectively.  The persistence diagrams computations are done by {\tt ripser} with tuning parameter {\tt n\_perm}=125.
}
\label{fig:demo rips pd resp}
\end{figure*}

\begin{figure*}[hbt!]
\centering
\includegraphics[trim=0 390 0 50,width=.95\linewidth]{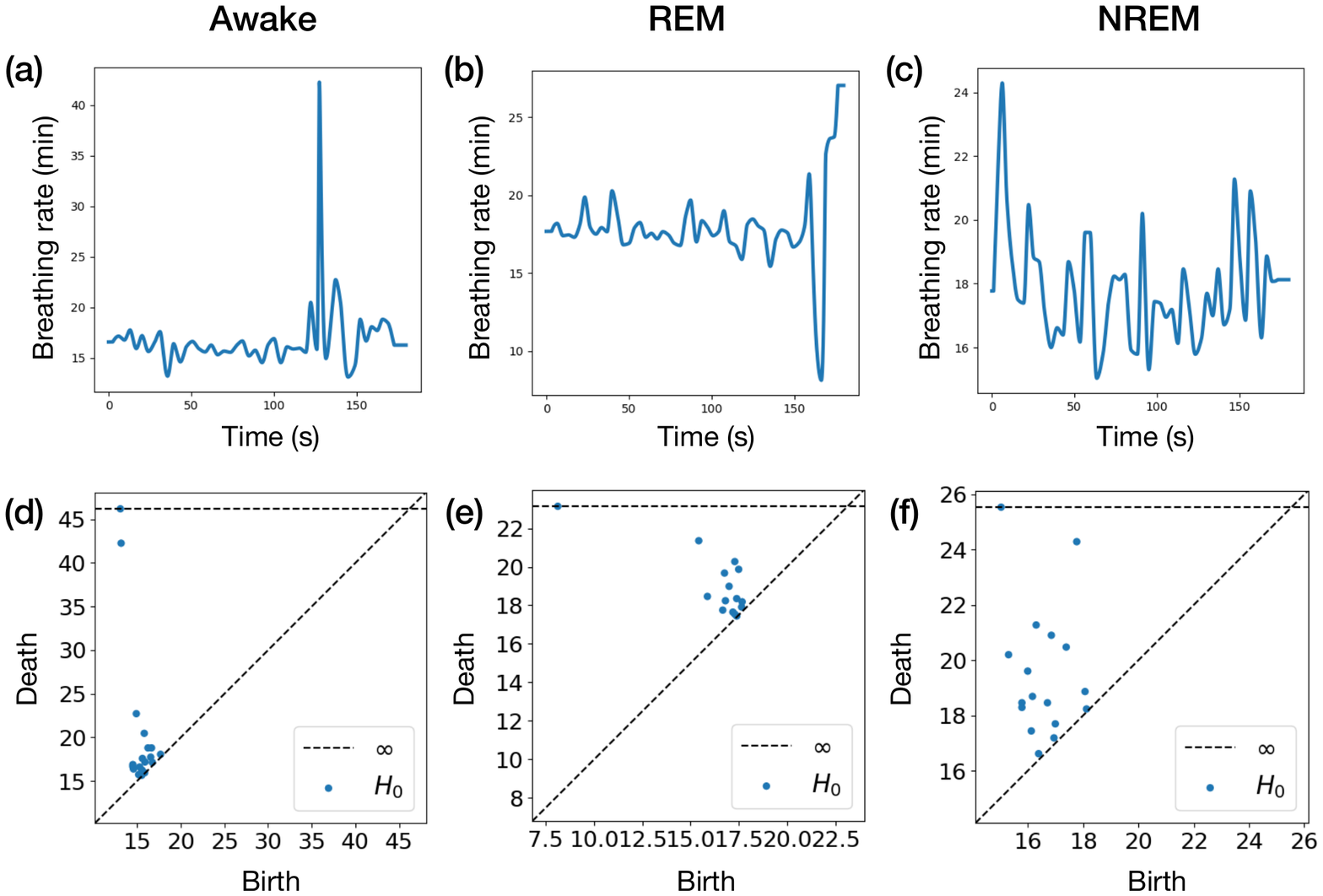}
\caption{Examples of instantaneous respiratory rate (IRR) and their persistence diagrams during different sleep stages. (a)-(c) IRR signals recorded during Wake, REM, and NREM sleep stages, respectively.  (d)-(f) persistence diagrams of the sublevel set filtrations of (a)-(c), respectively.}
\label{fig:demo rips pd irr}
\end{figure*}

\subsection{Features from persistence diagrams}

Although the space of PD is a metric space \cite{mileyko2011probability}, it can not be embedded into a Hilbert space \cite{bubenik2020embeddings,wagner2021nonembeddability}. Thus, applying machine learning algorithms directly on PDs is a challenging task (e.g. taking average over PDs involves a demanding theoretic development \cite{turner2014frechet}). Instead of using PDs directly, a common remedy is to convert a PD into finite dimensional vector or function, and this process is often called {\em summarizing PDs}. This {\em dimension reduction} approach is an active research area in TDA. There are a number of works in this direction, such as persistence landscape \cite{bubenik2015statistical}, persistence image \cite{adams2017persistence}, persistence indicator functions~\cite{rieck2019topological}, general functional summaries \cite{berry2018functional}, persistent entropy~\cite{atienza2020stability}, the Euler Characteristic Curve~\cite{richardson2014efficient}, lattice paths for persistence diagrams \cite{chung2021lattice}, persistence path\cite{chevyrev2018persistence}, persistence curve framework \cite{chung2022persistence}, and PersLay \cite{carriere2020perslay} and a recent survey article on this topic \cite{ali2022survey}). In this work, we consider {\em persistence statistics} (PS) \cite{chung2021persistent} and the recently developed Hermite coefficients on persistence curves \cite{chung2022persistence}.  PS are vectors consisting of elementary statistical summaries of birth and death values.  Persistence curves, on the other hand, are functional representation of PDs and Hermite coefficients of persistence curves are finite dimensional vector representation of the curves.  

To describe the summarization methods that we use in this article, we denote the PD by $D = \{(b_i,d_i)\}_{i\in \mathbb{N}}$, which is a collection of pairs of birth and death values.
Define the {\em midlife persistence} 
\[
D_m:=\{ (d+b)/2~;~ (b,d)\in D \}
\]
and {\em lifespan persistence} 
\[
D_l:=\{ d-b~;~ (b,d)\in D \}\,. 
\] 
Then, consider five basic summary statistics, including mean, standard deviation, skewness, kurtosis and entropy of $D_m$ and $D_l$; that is, mean($D_m$), std($D_m$), sk($D_m$), ku($D_m$), epy($D_m$), mean($D_l$), std($D_l$), sk($D_l$), ku($D_l$), epy($D_l$), where std means standard deviation, sk means skewness, ku means kurtosis, and epy means entropy.  Specifically, epy($D_m$) and epy($D_l$) are defined as
\begin{align}
    & \text{epy}(D_l) = -\sum_{(b,d)\in D} \frac{d-b}{L} \log\left(\frac{d-b}{L}\right), \\
    & \text{epy}(D_m) = -\sum_{(b,d)\in D} \frac{d+b}{M} \log\left(\frac{d-b}{M}\right),\nonumber
\end{align}
where $L = \sum_{(b,d)\in D} (d-b)$ and $M = \sum_{(b,d)\in D} (d+b)$.
In addition, we consider the the 1-norm of {\em Gaussian persistence curve} defined as
\begin{equation}
\label{equ:gpc}
    \| G_D \|_1 = \sum_{(b,d)\in D} \left[ (d-b) \Phi\left(\frac{d-b}{\sqrt{2}\sigma}\right) + \sqrt{2}\sigma\phi\left(\frac{d-b}{\sqrt{2}\sigma}\right) \right],
\end{equation}
where $\phi(x)$ is the probability density function of standard normal distribution and $\Phi(x)$ is the associated cumulative density function and $\sigma>0$ is a parameter.  In this work, $\sigma = 1$.  \eqref{equ:gpc} was proposed in \cite{chung2022gaussian} and applied in a scratch detection study \cite{juji2023}.  In summary, the PS of a PD is a 11-dimensional vector:
\begin{align*}
       \text{PS}(D):= &\,[ \text{mean}(D_m),~ \text{std}(D_m),~ \text{sk}(D_m),~ \text{ku}(D_m),~ \text{epy}(D_m),\\
        &\quad\text{mean}(D_l),~ \text{std}(D_l),~ \text{sk}(D_l),~ \text{ku}(D_l),~ \text{epy}(D_l),~ \| G_D \|_1]^\top.
\end{align*}

For the persistence curve, we consider the {\em lifespan entropy curve} \cite{chung2022persistence,atienza2020stability} defined as
\begin{equation}
    \text{le}(x) = -\sum_{(b,d)\in D} \frac{d-b}{L} \log\left(\frac{d-b}{L}\right) \mathbbm{1}_{[b, d)}(x),~x\in \mathbb{R},
\end{equation}
where $\mathbbm 1_{[b,d)}$ is an indicator function on the interval $[b,d)$.  To calculate $\text{le}(x)$, one common way is to discretize the given function.  Recently, an alternative approach based on Hermite function expansion is proposed in \cite{chung2022persistence}. In short, one writes persistence curves as a linear combination of Hermite functions, i.e. $\text{le}(x) = \sum_{n=0}^{\infty} \alpha_n h_n(x)$, where $h_n(x)$ is the n-th degree Hermite function.  According to Theorem III.2 in \cite{chung2022stable}, $\alpha_n$ can be found by the following formulas:
\begin{align}
		&\alpha_0 = \sum_{(b,d)\in D} \sqrt{2}\pi^{1/4} \psi(b,d) [ \Phi(d)-\Phi(b) ],\nonumber\\
  &\alpha_1 =  \sum_{(b,d)\in D} 2\pi^{1/4} \psi(b,d) [ \phi(b) - \phi(d)],\\
  &\alpha_{n+1} = \frac{\sqrt{2}}{\sqrt{n+1}}\left[\sum_{(b,d)\in D} \psi(b,d)  ( h_n(b) - h_n(d) )\right]+ \frac{n}{\sqrt{n(n+1)}} \alpha_{n-1}, \nonumber
\end{align}
for $n \geq 1$, where $\psi(b,d) = -\frac{d-b}{L} \log(\frac{d-b}{L})$.  In this work, we use $\alpha_n$ for $n=0,1,2,\ldots,14$ as features.  In summary, in this work, the Hermite function expansion of persistence curve (HEPC) is a 15-dimensional vector:
\begin{equation}
    \text{HEPC}(D):= [\alpha_0, \alpha_1,\ldots,\alpha_{14} ]^\top.
\end{equation}

\section{Automatic sleep stage annotation}
\label{sec:sleep stages}

We apply the developed TDA-based summaries of time series to the challenging automatic sleep stage annotation problem when we only have the airflow signal. In addition to demonstrating the performance of these TDA-based features, we also implement existing commonly used features and provide a comparison.  There are five sleep stages: Wake, Rapid Eye Movement (REM), Non-REM 1 (N1), Non-REM 2 (N2), and Non-REM 3 (N3).  We focus on establishing an automatic sleep stage scoring of wake, REM and NREM.

\subsection{Data}
The proposed method is evaluated in a real-world dataset that consists of 30 overnight polysomnogram recordings recorded at the sleep center in Chang Gung Memorial Hospital (CGMH), Linkou, Taoyuan, Taiwan. All subjects were suspected of sleep apnea and referred to the sleep lab, and they were diagnosed to be free of sleep apnea; that is, the apnea-hypopnea index (AHI) is less than 5. The age is 45.2$\pm$15.0 years old with the range 19 to 80 years old and the body mass index is $24.2\pm 3.5$. The total sleep time is 6.19$\pm$0.97 hours, the ratio of wake (REM, N1, N2 and N3 respectively) stage is $17.3\%\pm 11.8\%$ ($14.0\%\pm 6.5\%$, $10.8\%\pm 5.6\%$,  $49.8\%\pm 10.3\%$ and  $8.1\%\pm 9.8\%$ respectively), and the AHI is $3.13\pm 1.54$. Four cases out of 30 do not have N3 sleep stages. Twelves subjects are male. All recordings were acquired on the Alice 5 data acquisition system (Philips Respironics, Murrysville, PA). Each recording is at least 5 hours long. The sleep stages were scored by two licensed, experienced sleep technologists abiding by the AASM criteria \cite{AASM2020}, and a consensus was reached. 
We focus on the airflow signal that is recorded from the nasal cannula. The airflow signal is sampled at 100 Hz.

\subsection{Preprocessing}

Suppose the given airflow signal $f_0$ is sampled at $\gamma$Hz, where $\gamma\in \mathbb{N}$. Suppose the length is $T\in \mathbb{N}$ seconds; that is, $f_0\in \mathbb{R}^{\gamma T}$. We apply the NeuroKit package \cite{Makowski2021neurokit} with the default setup (linear detrending followed by a fifth order $2$Hz low-pass IIR Butterworth filter) to obtain the exhalation onset of each breathing cycle. Denote $t_i$ as the timestamp of the inspiration of the $i$-th breathing cycle. Suppose there are $N$ detected breathing cycles. Define the instantaneous respiratory rate (IRR) via the monotone cubic spline of the nonuniform samples
\[
\left\{\left(t_i, \frac{60}{t_i-t_{i-1}}\right)\right\}_{i=2,\ldots, N},
\]
and sample the resulting function at $4$Hz, denoted as $r\in \mathbb{R}^{4T}$. The unit of $r(i)$ is cycles per minute. Note that the higher the $r(i)$ is, the faster the breathing is. Also, the higher fluctuation of $r$ from time to time, the higher variability the breathing is. We follow the convention and divide the airflow signal, and hence the IRR signal, into non-overlapping 30 seconds epochs. The division coincides with that for the sleep stage annotation.

Next, we evaluate the signal quality of the airflow signal for each epoch. For the $i$-th epoch, where $i>5$, denote the airflow signal over the $i-5$th to $i$th epochs as $f_i$. Apply an 3rd order Butterworth bandpass filter with the spectral band $[0.1,\ 0.75]$ Hz to $f_i$, and use the same notation to denote the resulting signal.
We evaluate the signal quality of $f_i$ by the spectrum-based signal quality index (SQI) \cite{birrenkott2015respiratory}, which is defined as
\[
SQI(i) = \frac{\sum_{l=\xi_0-2}^{\xi_0+2} |\widehat{f}_i(l)|^2}{\sum_l |\widehat{f}_i(l)|^2}\,,
\]
where $\widehat{f}_i$ is the discrete Fourier transform of $f_i$, and $\xi_0$ is the detected spectral peak of $f_i$ over $[0.1,\ 0.75]$ Hz.  Ideally, when the airfow signal has a high quality, SQI should be high since there is a peak in the spectral domain due to the oscillating pattern. On the other hand, a low SQI indicates a questionable signal quality.

\subsection{Features and learning model}
To design features for the $i$th epoch, where $i>5$, we consider the airflow signal and the associated IRR signal over the current and five previous epochs, denoted as $f_i\in \mathbb{R}^{\gamma_{i}\times 180}$ and $r_{i}\in \mathbb{R}^{4\times 180}$ respectively, where $180$ corresponds to the length of 6 epochs. There are four PDs we consider for the airflow signal and the associated IRR signal, including
$D_0^{\texttt{Sub}}(f_i)$, $D_0^{\texttt{Rip}}(f_i)$, $D_1^{\texttt{Rip}}(f_i)$, and $D_0^{\texttt{Sub}}(r_i)$. The topological features we use are PS($D_0^{\texttt{Sub}}(r_i)$), HEPC($D_0^{\texttt{Sub}}(r_i)$), HEPC($D_0^{\texttt{Rip}}(f_i)$), PS($D_1^{\texttt{Rip}}(r_i)$), PS($D_0^{\texttt{Sub}}(f_i)$), HEPC($D_0^{\texttt{Sub}}(f_i)$).
In our implementation, the Rips complex filtration of $f$ is calculated with the delay parameter $d=3$ and $\tau=100$ in the Takens' embedding.

We also consider classic non-TDA features; 
particularly, we implement $33$ features proposed in \cite{tataraidze2015sleep}, including features in the time and frequency domains, entropies, signal quality and dynamic time and frequency warping. Particularly, we follow the details described in \cite[Table III]{long2013sleep} and \cite{long2014analyzing} to implement the dynamic time and frequency warping and the depth-based and volume-based features respectively, and apply the NeuroKit package \cite{Makowski2021neurokit} to generate spectral domain and entropy features.
The implementation of features proposed in \cite{tataraidze2015sleep} also serves as a comparison with the existing state-of-the-art work.

\subsection{Learning and evaluation}

We choose XGBoost (eXtreme Gradient Boosting) \cite{xgboost} as our classifier, which is a popular open-source implementation of gradient boosting framework designed for efficiency and scalability. The main objective is to learn the class labels, namely, wake, REM and NREM. We choose the following XGBoost setting: compared with the default setting, the learning rate (step size shrinkage used in update) is slightly reduced to 0.07, the maximum depth of a tree is slightly increased to $5$, the subsample ratio of the training instances is reduced to 0.2, and the subsample ratio of columns when constructing each tree is reduced to 0.5.

We consider three sets of features. The first set contains the non-TDA features proposed in \cite{tataraidze2015sleep}, denoted as \texttt{ntda}; the second set contains the TDA features; denoted as \texttt{tda}; the third set contains both non-TDA and TDA features, denoted as \texttt{all}. 
We further consider taking the SQI into account during the training process as respiratory signals are sensitive to body motion artifacts \cite{long2014analyzing}. Specifically, involving low quality signal would substantially downgrade the model training. Thus, if the median of SQI over an epoch is lower than $0.25$, the  epoch is removed in the training process. This value is chosen to exclude visually identifiable erroneous airflow signals and not tuned to increase the overall performance of the trained model.

To evaluate the performance of the trained model with the chosen features, we consider the leave-one-subject out cross validation (LOSOCV) scheme. This cross validation is done by removing the data associated with one subject from the training process, and predict the class labels given the features on that subject, and repeat this process to cycle through all the subjects. This scheme mimics the real world scenario to ensure that the chosen model generalizes well to new subjects. We report the average performance of our LOSOCV in the next section. 
To evaluate the classification performance, we consider sensitivity for each class, accuracy, balanced accuracy, and Cohen's kappa score, where the balanced accuracy is defined as the average of sensitivities of different classes. 
Since Cohen's kappa is less sensitive to imbalance of the class labels, we view it as a more important metric than accuracy in our study. The one-sided Wilcoxon signed rank test, with the significance level set to $p=0.05$, is applied to compare the classification performance between two methods.

\subsection{Results}

During the training process, we remove median 27.5\% and median absolute deviation 13.2\%, or  27.5\%$\pm$13.2\% (11.3\%$\pm$9.3\%, 9.9\%$\pm$12.2\%, 2.1\%$\pm$5.1\% and 0\%$\pm$0.4\% respectively), epochs labeled as wake (REM, N1, N2 and N3 respectively) with the SQI threshold set to $0.25$. 

\begin{figure}[hbt!]
\centering
\includegraphics[trim=0 0 0 0,width=1\linewidth]{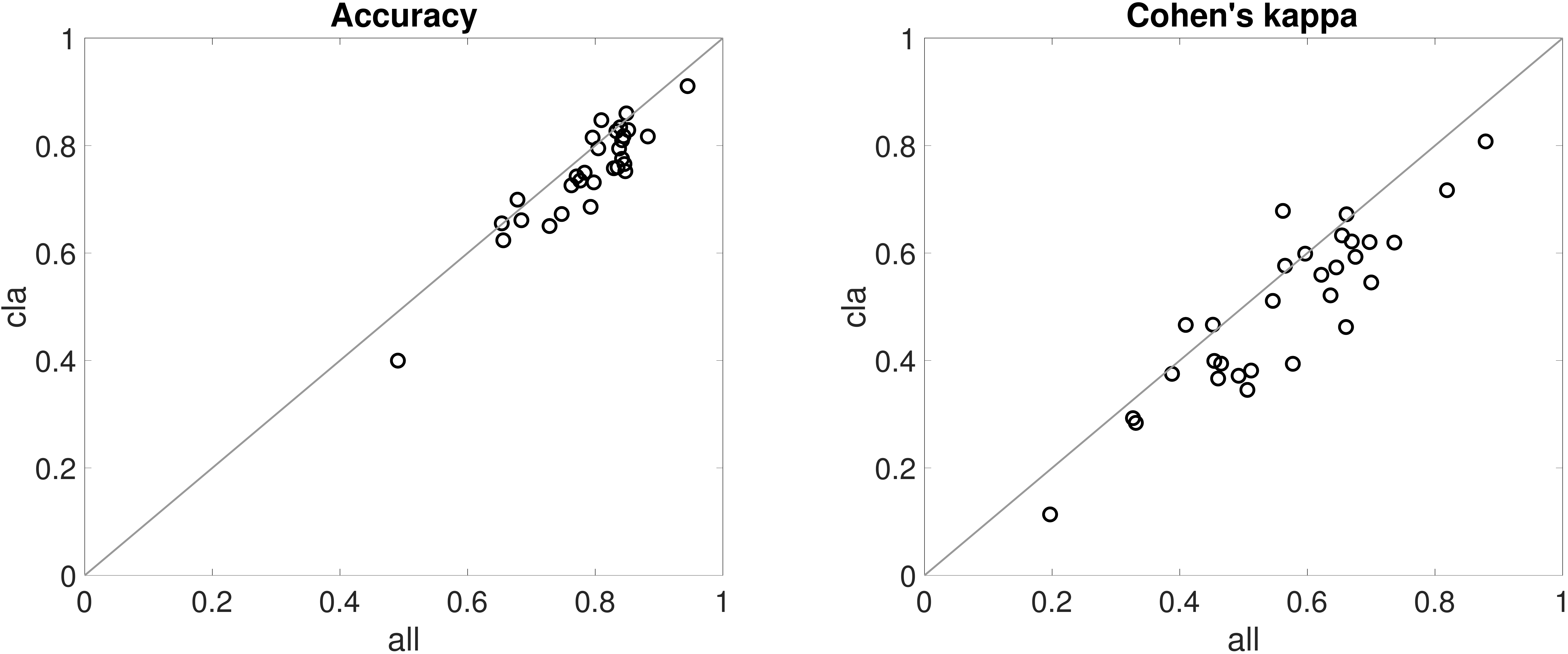}
\caption{A comparison of classification performance, in terms of accuracy (Left) and Cohen's kappa (right), using the proposed feature set \texttt{(all)} and the non-TDA feature set \texttt{(cla)}.}
\label{fig:comparison}
\end{figure}

\begin{figure}[hbt!]
\centering
\includegraphics[trim=0 0 0 0,width=1\linewidth]{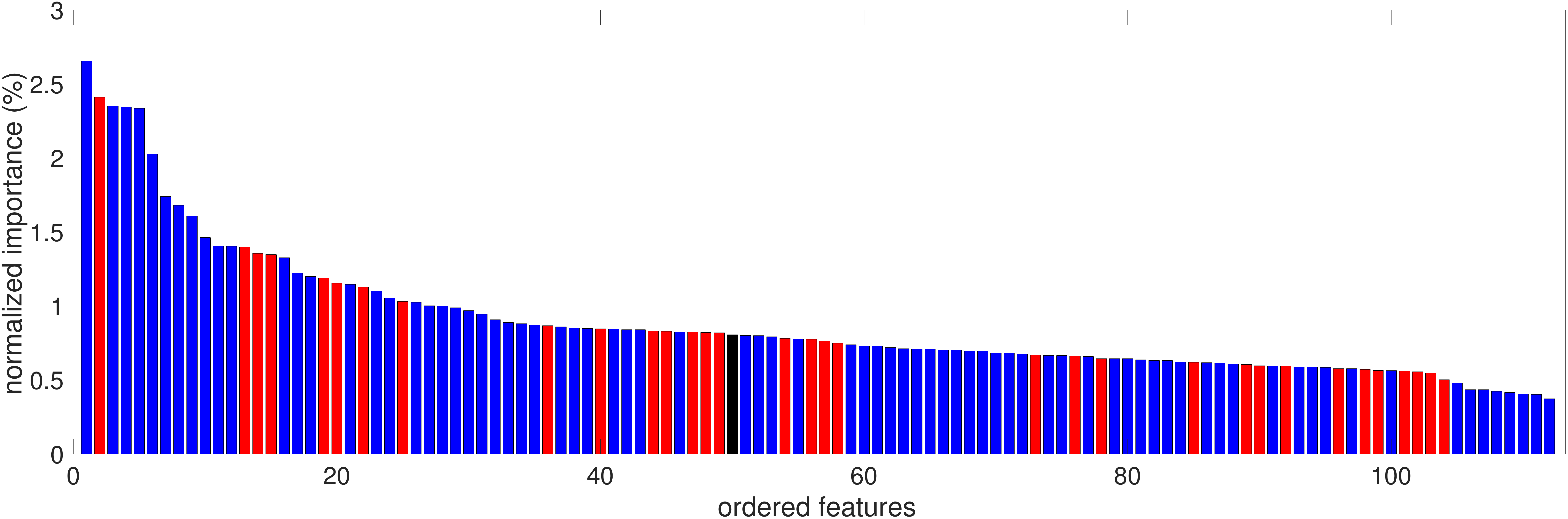}
\caption{The importance of each feature in the proposed feature set \texttt{all}. Features in the feature set \texttt{cla} are colored in red, features in the feature set \texttt{tda} are colored in blue, and RQI is colored in black.}
\label{fig:importance}
\end{figure}

The three-label classification results of wake, REM and NREM (N1, N2 and N3) stages with different features sets are shown in Table \ref{tab:3 classes}. Averages over 30 subjects with one standard deviation are shown. The class weights for wake, REM, NREM are set to 4, 4, and 1 respectively to handle the imbalanced data. Overall, the model with the feature set \texttt{all} performs better than that with the feature set \texttt{cla} considered in \cite{tataraidze2015sleep} with statistical significance (p-value $0.05$ in terms of Cohen's kappa, and $0.02$ in terms of overall accuracy). See Table \ref{tab:conf_table} for the confusion matrices from three different feature sets. Specifically, the TDA features help improve the sensitivities of detecting wake and REM sleep stages, and hence the overall accuracy and Cohen's kappa. Moreover, in Figure \ref{fig:comparison}, we can clearly see that in most cases both overall accuracy and Cohen's kappa are higher (i.e., points below the 1-1 line) if the feature set $\texttt{all}$ is used.  To check if the TDA features are useful, we plot the normalized averaged information gain as the feature importance calculated by xgboost over the LOSOCV in Figure \ref{fig:importance}, where features in the feature set \texttt{cla} are colored in red, features in the feature set \texttt{tda} are colored in blue, and RQI is colored in black. We can see that over the top 20 most important features and 14 of them are TDA features, and 6 of them are non-TDA features. This suggests that TDA features contribute to the classifier. 

\begin{table}[h]
\centering
(a)
\begin{tabular}{|c|c|c|c|c|}
\hline
\multicolumn{2}{|c|}{} &\multicolumn{3}{c|}{\textbf{Prediction}}\\
\cline{3-5}
\multicolumn{2}{|c|}{\texttt{all} feature set}  & \textbf{wake} & \textbf{REM} & \textbf{NREM} \\
\hline
\multirow{3}{*}{\begin{sideways} \textbf{True} \end{sideways}}&\textbf{wake} & 62.1 & 5.5 & 18.9 \\
\cline{2-5}
&\textbf{REM} & 8.2 & 58.8 & 14.3 \\
\cline{2-5}
&\textbf{NREM} & 45.9 & 35.4 & 342.9 \\
\hline
\end{tabular}
\vspace{10pt}\\
(b)
\begin{tabular}{|c|c|c|c|c|}
\hline
\multicolumn{2}{|c|}{} &\multicolumn{3}{c|}{\textbf{Prediction}}\\
\cline{3-5}
\multicolumn{2}{|c|}{\texttt{tda} feature set}  & \textbf{wake} & \textbf{REM} & \textbf{NREM} \\
\hline
\multirow{3}{*}{\begin{sideways} \textbf{True} \end{sideways}}&\textbf{wake} & 61.4 & 6.4 & 18.8 \\
\cline{2-5}
&\textbf{REM} & 9.3 & 53.0 & 19.1 \\
\cline{2-5}
&\textbf{NREM} & 58.0 & 42.8 & 323.3 \\
\hline
\end{tabular}
\vspace{10pt}\\
(c)
\begin{tabular}{|c|c|c|c|c|}
\hline
\multicolumn{2}{|c|}{} &\multicolumn{3}{c|}{\textbf{Prediction}}\\
\cline{3-5}
\multicolumn{2}{|c|}{\texttt{cla} feature set}  & \textbf{wake} & \textbf{REM} & \textbf{NREM} \\
\hline
\multirow{3}{*}{\begin{sideways} \textbf{True} \end{sideways}}&\textbf{wake} & 55.1 & 12.7 & 18.8 \\
\cline{2-5}
&\textbf{REM} & 10.8 & 52.7 & 17.9 \\
\cline{2-5}
&\textbf{NREM} & 45.3 & 45.7 & 333.2 \\
\hline
\end{tabular}
\caption{The average confusion matrix from the leave-out-subject out cross validation. From (a) to (c), the \texttt{all}, \texttt{tda} and \texttt{cla} feature sets are used separately.}
\label{tab:conf_table}
\end{table}

We shall mention that when the epochs with low signal quality are not excluded from the training dataset, the results are all consistently worse with statistical significance. See Table \ref{tab:5 classes nrqi} for details. This result is consistent with the fact that if a model is trained on noisy data, it tends to also fit the noise in the data, which leads to a high variance in the model and hence a poor performance on new, unseen data.

\begin{table*}
\small
	\centering
	\begin{tabular}{|c|c|c|c|c|c|c|}
		\hline
		&wake & REM & NREM & Acc & Balanced & $\kappa$ \\ \hline\hline
		\texttt{all} & 0.710 $\pm$ 0.224 & 0.737 $\pm$ 0.209 & 0.813 $\pm$ 0.110 & 0.788 $\pm$ 0.087 & 0.753 $\pm$ 0.111 & 0.563 $\pm$ 0.150 \\ \hline
		\texttt{tda} &0.694 $\pm$ 0.212 & 0.681 $\pm$ 0.220 & 0.765 $\pm$ 0.132 & 0.742 $\pm$ 0.100 & 0.713 $\pm$ 0.106 & 0.489 $\pm$ 0.147 \\ \hline
		\texttt{ntda} & 0.653 $\pm$ 0.184 & 0.667 $\pm$ 0.221 & 0.794 $\pm$ 0.108 & 0.750 $\pm$ 0.096 & 0.704 $\pm$ 0.108 & 0.499 $\pm$ 0.151
 \\ \hline
	\end{tabular}
	\caption{Leave one subject out cross validation results of learning and testing three sleep stages annotation with different feature sets. The second to the fourth columns represent the sensitivity for Wake, REM, and NREM (N1, N2 and N3) class, respectively.  Acc represents the overall accuracy, Balanced represents the balanaced accuracy, and $\kappa$ represents the Cohen's kappa.} 
	\label{tab:3 classes}
\end{table*}

\begin{table*}
\small
	\centering
	\begin{tabular}{|c|c|c|c|c|c|c|}
		\hline
		&wake & REM & NREM & Acc & Balanced & $\kappa$ \\ \hline\hline
		\texttt{all} &  0.736 $\pm$ 0.198 & 0.590 $\pm$ 0.299 & 0.734 $\pm$ 0.160 & 0.709 $\pm$ 0.108 & 0.687 $\pm$ 0.116 & 0.458 $\pm$ 0.145 \\ \hline
		\texttt{tda} &0.764 $\pm$ 0.190 & 0.579 $\pm$ 0.290 & 0.666 $\pm$ 0.202 & 0.660 $\pm$ 0.151 & 0.670 $\pm$ 0.129 & 0.411 $\pm$ 0.164 \\ \hline   
		\texttt{ntda} & 0.696 $\pm$ 0.204 & 0.560 $\pm$ 0.249 & 0.730 $\pm$ 0.154 & 0.693 $\pm$ 0.108 & 0.662 $\pm$ 0.103 & 0.430 $\pm$ 0.135 \\ \hline
	\end{tabular}
	\caption{Results without removing low quality epochs from the training dataset during the leave one subject out cross validation. The second to the fourth columns represent the sensitivity for wake, REM, and NREM (N1, N2 and N3) sleep stage, respectively.  Acc represents the overall accuracy, Balanced represents the balanced accuracy, and $\kappa$ represents the Cohen's kappa.}
	\label{tab:5 classes nrqi}
\end{table*}

\section{Discussion and Conclusion} \label{sec6}

In this paper, we show the great potential of utilizing the quantified BPV from an airflow signal by TDA  to design an accurate automatic sleep stage scoring system. From the physiological standpoint, sleep encompasses not only brain activity, but also affects every facet of the body, leading to variations in the associated physiological signals \cite{kryger2017principles}. As summarized in Introduction, analyzing non-EEG signals has emerged as a viable technique for determining sleep stages. In recent years, this approach has gained considerable traction, particularly in the field of homecare digital medicine. The ability to monitor sleep dynamics outside of a clinical setting has the potential to transform the way we understand sleep and its impact on health. Furthermore, it enables patients to track their sleep patterns from the comfort of their homes, promoting personalized care and facilitating early detection of sleep disorders.

In this paper, we focus on designing an automatic sleep stage scoring system using only the airflow signal. In general, the respiratory signals comprise a wealth of information other than sleep pertaining to the physiological dynamics that researchers and clinicians can harness to inform clinical decision-making \cite{hlastala2001physiology}.
The control of breathing and respiration is complicated, which includes but exclusively the neural activity from the respiratory center, the arterial chemoreceptors, and lung sensory receptors and mechanics \cite{hlastala2001physiology}. Depending on where the respiratory signal is collected, the signal further depends on the specific anatomical structure and mechanics, electrophysiological dynamics and others. For example, the airflow signal we focus on in this paper encodes rich information about the orolarynpgeal anatomical structure. These factors are integrated in a complicated way leading to BPV, whose origin can be reasoned similarly to that of heart rate variability (HRV) \cite{berntson1997heart}.  BPV can be roughly classified into two main categories from the signal processing perspective \cite{van1991respiratory,engoren1998approximate,benchetrit2000breathing}. The first of these is respiratory rate variability (RRV). RRV describes fluctuations in the duration of breathing cycles that is encoded in the IRR signal we study in this paper.
The second is the oscillatory pattern variability, which refers to changes in the breathing cycle's pattern. A typical example of oscillatory pattern variability is the inspiration/expiration time ratio (IER), which has been widely explored \cite{burtt1921inspiration}. The results shown in this paper suggest the potential of applying TDA to study BPV from respiratory signals, and apply it to study other problems. We will explore this possibility in our future work.

To our knowledge, there are few published papers that studied automated sleep stage scoring solely based on the respiratory signal, which we summarize here. 
The work \cite{sloboda2011simple} considered 9 features in the time and frequency domains derived from the nasal airflow signal and abdominal respiratory effort signal recorded from inductance plethysmography. The learner is the naive Bayes classifier. The analysis shows about 70\% accuracy from 16 healthy subjects.
In \cite{long2014analyzing}, the authors designed a set of 12 novel features reflecting respiratory depth and volume from thoracic respiratory effort signal recorded from inductance plethysmography and train a linear discriminant classifier with these 12 features in addition to 14 traditional ones. The signal calibration by body movements detected by dynamic time warping measures and subject-specific normalization are considered to improve the overall performance. This system achieved average accuracy of 76.2\%$\pm$7.9\% and Cohen's kappa coefficient of $0.45\pm 0.15$ with recordings from 48 health subjects using a 10-fold cross validation on the subject level. 
In \cite{long2014measuring}, 27 features reflecting respiratory morphology from thoracic respiratory effort signal recorded from inductance plethysmography are considered to train a linear discriminant classifier. The subject-specific normalization is considered to improve the overall performance.
This system achieved average accuracy of 77.1\%$\pm$7.6\% and Cohen's kappa coefficient of $0.48\pm 0.17$ with 48 recordings from health subjects.
In \cite{tataraidze2015sleep}, thirty-three non-TDA features are extracted from respiratory inductive plethysmography signal recorded from the thoracic belt, and the bagging classifier \cite{breiman1996bagging} is applied to automatically score wake, REM and NREM. The accuracy is $77.85\% \pm 6.63\%$ and Cohen's kappa is $0.59\pm 0.11$ when the algorithm is applied to 29 subjects subjects without sleep-related breathing disorders via the LOSOCV scheme. The subject-specific normalization is considered to improve the overall performance.
Some postprocessing heuristics are considered, including scoring all epochs in the first 20 minutes as wakefulness, score an epoch as previous stage if it did not belong to one of the nearest stages, all REM epochs during the first 60 minutes of records are scored as previous stage, and if an interval between REM epochs was less than 15 minutes, all epoch included in the interval were scored as REM.With these heuristics, the accuracy is increased to $80.38 \%\pm 8.32\%$ and the Cohen's kappa to $0.65 \pm 0.13$. 
In \cite{yang2016sleep}, two RRV features, the mean respiratory rate and the max absolute differences of breath intervals, are extracted from the oro-nasal airflow signal. A two-stage system with postprocessing rules is applied as the learner of wake, REM and NREM. The first stage detects wake and REM, and the second stage separates REM and Awake based on the outputs from the first layer. The postprocessing rules are based on sleep theory --- If the detected wake and REM epochs in the first layer occur in the first 60 minutes, these epochs were assigned as wake, and if the duration of a block of continuous detected wake and REM epochs in the first layer was less than 5 minutes, these epochs were assigned as Awake. 
This system achieved average accuracy of 74.00\%$\pm$5.30\% and Cohen's kappa coefficient of $0.49\pm 0.08$ with 21 recordings from health subjects. 

Overall, our result cannot be fairly compared with these reported results due to some setup variations. The setup in \cite{tataraidze2015sleep} is closest to ours, and we consider their non-TDA features in this work for a comparison. While applying these non-TDA features in this work, we cannot achieve the same accuracy and Cohen's kappa as those reported in \cite{tataraidze2015sleep}, but it does provide decent information about sleep stages. Moreover, when we combine these non-TDA features with the proposed TDA features, the overall performance improves. We shall mention that the TDA features of the airflow signal is close to, but different from, the features considered in \cite{long2014measuring} in the sense of analyzing the morphology. Note that in our system, we do not enforce any postprocessing rules to keep the generalizability of the developed model. Also, while the overall accuracy and Cohen's kappa are lower than our results, considering only two RRV features are used, the overall performance of \cite{yang2016sleep} is impressive.

The current work has some limitations that open up possibilities for interesting future research. Firstly, the sample size is limited to only $30$ cases, and all subjects in the study do not have sleep apnea. Therefore, it would be beneficial to expand the dataset to include a larger and more diverse sample to better evaluate the feasibility and effectiveness of the proposed model and features. Specifically, exploring the possibility of applying the proposed algorithm to estimate sleep stages in patients with sleep apnea is an important avenue for future research. This will help generalize of the applicability of the proposed algorithm. From a technical standpoint, there are numerous intriguing TDA features that we could investigate while working with the Rips complex filtration. However, the time required for extracting TDA features remains a challenge. Therefore, an in-depth exploration of computationally efficient algorithms is necessary to facilitate real-time implementation. This might not only lead to better results, but enable us to leverage TDA features in a more effective and efficient manner. While we have shown the potential of TDA features and its geometric picture is clear, its interpretability in the sense of physiology is not clear. A serious research establishing the relationship between physiology and these features would be a fascinating topic for future research.  
Another avenue for future research is exploring the possibility of applying the proposed TDA-based BPV quantification technique to other respiratory channels. Doing so could lead to a deeper understanding of how the proposed method behaves across different respiratory channels, and provide insights into the generalizability of the approach. 
We will address these questions in our future work.

\bibliographystyle{plain} 
\bibliography{BPV_TDA.bib}       %

\end{document}